**Noncontact Chemical and Morphological Profiling in Liquid via Fiber-Based Micro-FTIR**

Ching-Ping Chang,[1,2] Jaehyeon Kim,[1,2] Guanhua Qiao,[1,2] Jonathan Matsuura,[2,4] Jimai Chen,[1,2] Lalith Krishna Samanth Bonagiri,[1,2] Yiran Zhao,[1,2] and Yingjie Zhang[1,2,3,a)]

[1]Department of Materials Science and Engineering, University of Illinois, Urbana, Illinois 61801, USA

[2]Materials Research Laboratory, University of Illinois, Urbana, Illinois 61801, USA

[3]Beckman Institute for Advanced Science and Technology, University of Illinois, Urbana, Illinois 61801, USA

[4]Department of Chemistry, University of Illinois, Urbana, Illinois 61801, USA

a) Correspondence to: yjz@illinois.edu

## ABSTRACT

Liquid-phase mid-infrared (MIR) imaging is challenging due to the strong MIR light absorption of most liquid media. Existing liquid-phase MIR microscopy approaches overcame the background absorption using predominantly either ultrathin liquid cells or attenuated total reflection (ATR) setups. However, these designs introduce additional challenges, such as sample confinement, limited probe volume, probe perturbation, and/or lack of quantification capabilities. In this work, a fiber-based Fourier-transform infrared (FTIR) microspectroscopy method was developed to overcome the technical barriers of liquid-phase MIR characterization. By integrating a tapered fiber microprobe, a transmission-based optical setup, and a scanning stage, we achieved noncontact chemical and morphological profiling in liquid. The micro-FTIR capabilities were benchmarked by measuring a micropatterned photoresist substrate in water, obtaining quantitative spatial chemical distribution and morphological profiles with ~35 μm lateral resolution and nanometer-scale vertical detection limit. This noncontact, quantitative, liquid-phase micro-FTIR technique is promising for non-perturbative characterization of heterogeneous structures in liquids.

## TOC Graphic

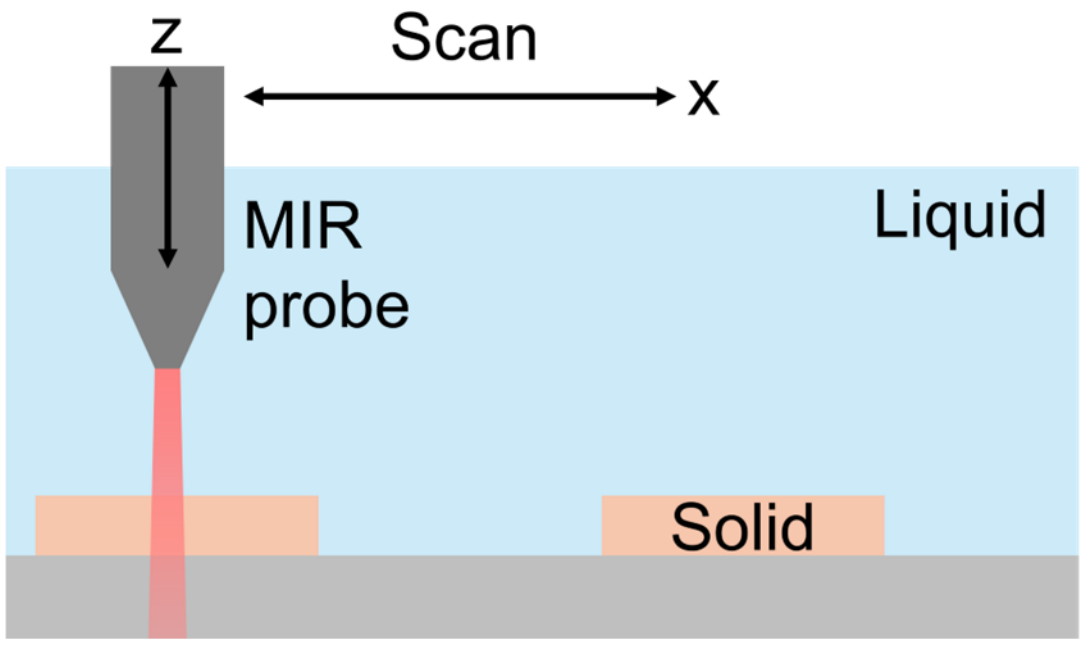

## 1. INTRODUCTION

Many chemical, electrochemical, and biological systems rely on liquids as essential functional media.[1–3] In these systems, the core underlying interaction, reaction, and transport processes require the active participation of both the immersed solid (electrode, macromolecule, etc.) and the surrounding liquid. Microscopic heterogeneities at the solid–liquid interfaces, on both the solid and liquid sides, are oftentimes critical for the overall function of the system. In electrochemical systems, for example, the electrodes—microparticle electrodes for batteries, porous gas diffusion electrodes for fuel cells and other electrocatalytic systems, etc.—often have multiscale structural heterogeneities with varying electrochemical activities.[4–6] The near-surface electrolytes in these systems are also spatially nonuniform, since mass transport results in concentration gradients within tens to hundreds of microns from the electrode surface; the spatial concentration profiles of reactants, intermediates, and products within this diffusion layer have significant impacts on the overall electrochemical reaction kinetics.[6–8] Despite their distinct functionalities, microscopic heterogeneities across phase boundaries also universally occur in biological building blocks, including the extracellular matrix, cell, and subcellular structures.[9–11] These spatial heterogeneities, regulated by membrane compartmentalization and liquid–liquid phase separation,[12,13] strongly influence molecular transport, gene expression, cell signaling, cellular stress response, and many other biological functions.[14–16]

To resolve the spatial morphology and chemical distribution of solid–liquid interfaces, numerous imaging techniques have been utilized, including electron microscopy, optical microscopy, and scanning probe microscopy.[17–19] These techniques are excellent in providing structural and morphological information of solid surfaces with micrometer to subnanometer spatial resolution. However, most of them provide little chemical information of the near-surface liquid species, which can extend up to hundreds of microns away from the solid surface. Vibrational spectromicroscopy, including Raman and MIR, offers label-free chemical imaging by spatially probing the molecular vibrational modes.[20,21] While Raman and MIR methods are oftentimes complementary, MIR is usually preferred for precise quantification purposes, as the MIR absorption directly obeys the Beer-Lambert law.[22] In contrast, Raman intensity depends not only on the concentration of the probed species but also on laser power, focal volume, collection efficiency, optical alignment, refractive index, polarization, detector response, and sometimes fluorescence; these factors can make absolute quantification less robust.[23–25]

A key challenge for liquid-phase MIR characterization is the strong background absorption of the liquid.[26–28] This limitation is particularly severe for aqueous systems, as water exhibits intense and broad absorption bands across the MIR spectral range.[26,29,30] To mitigate this challenge, several approaches have been explored (Figure 1). One approach combines thin liquid cells (<10 μm) with strong synchrotron MIR radiation (Figure 1a), enabling noncontact spectromicroscopy with a spatial resolution down to ~2–10 μm.[31–34] However, the liquid-cell method poses stringent requirements on the sample size (smaller than the cell path length).[35] Additionally, conventional MIR microscopy uses large numerical aperture objective lens to achieve high imaging resolution,[36] resulting in highly noncollimated beam that complicates the quantification of the signal.[22] Alternatively, micro-ATR-FTIR uses an evanescent field to mitigate bulk absorption and enables imaging with micron-scale spatial resolution using a focal plane array (FPA) detector (Figure 1b).[37–40] However, the probing depth is limited to ~1 μm surrounding the ATR crystal, and quantitative analysis of ATR measurements can be challenging due to uncertainties in the exact path length.[22,41–43] Another approach to enable liquid-phase MIR imaging is to combine atomic force microscopy (AFM) with MIR spectroscopy. Depending on the detection algorithm, these nano-IR methods include scattering-type scanning near-field optical microscopy (s-SNOM), photothermal induced resonance (PTIR), and peak force infrared (PFIR) microscopy.[44–50] Recently, ultrathin membranes have been combined with s-SNOM to mitigate strong water absorption by separating the AFM tip from the aqueous environment (Figure 1c).[51–53] However, the probing depth of these existing nano-IR techniques is limited to ~100 nm.[51,52] In addition, quantitative interpretation remains challenging because the detected signals are strongly influenced by tip geometry, near-field scattering, photothermal expansion, and/or tip–sample coupling.[48,54–56] Furthermore, the direct or intermittent tip–sample contact during these nano-IR measurements can perturb delicate samples and/or the nearby liquid environment. Overall, existing liquid-phase MIR micro- and nano-spectroscopy techniques face limitations in probing depth, quantification capability, and/or sample perturbation.

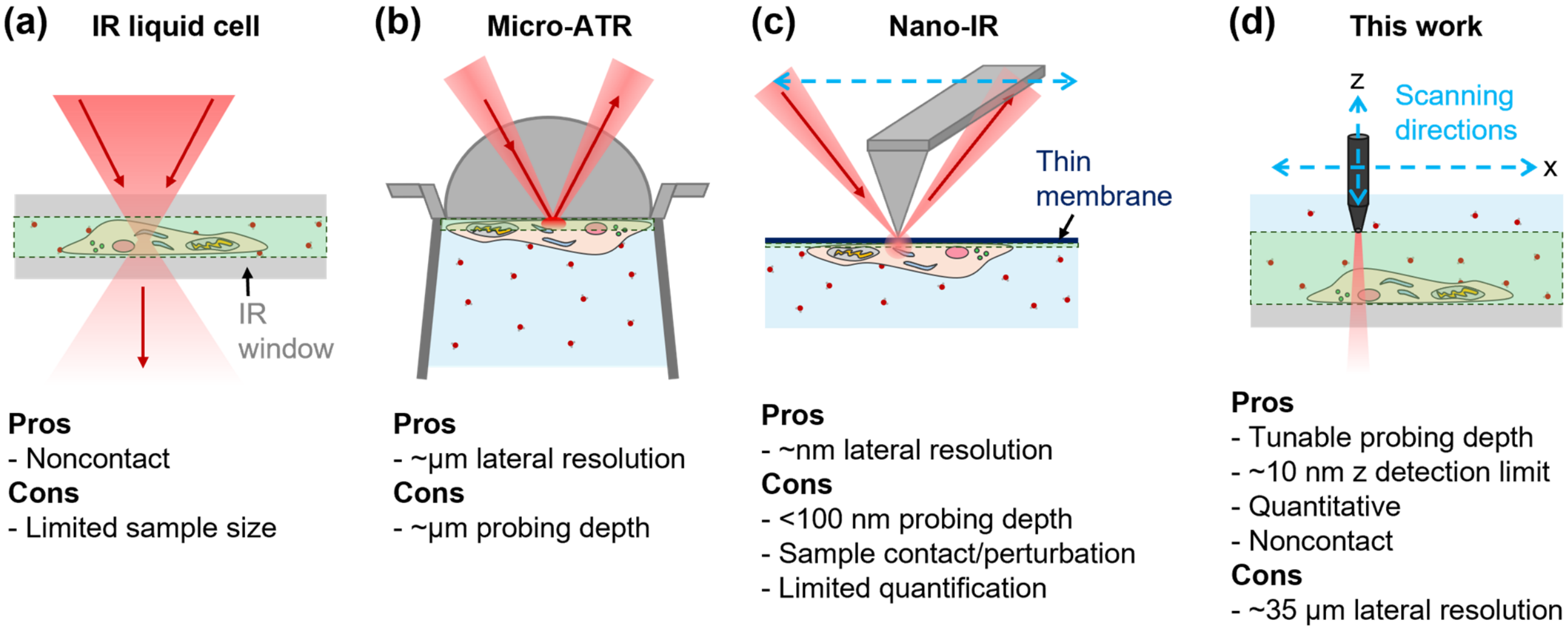

**Figure 1. Comparison of different MIR spectromicroscopy methods for liquid-phase chemical and/or morphological mapping.** The green dashed boxes mark the active probing volume of each method.

To overcome the challenges of existing liquid-phase MIR measurements, our lab recently demonstrated an endoscopic FTIR (endo-FTIR) method, where we used a fiber probe to achieve variable-path-length FTIR characterization inside liquids.[57,58] By controlling the path length ($z$) with ~1 μm steps and down to a minimum of ~3 μm, we achieved precise quantification of the liquid absorption coefficient ($\alpha$) up to ~$3\times10^3$ cm$^{-1}$ over a large portion of the MIR spectral range. Nevertheless, two major limitations still persisted: 1) in aqueous solutions, MIR transmittance within ~3100–3700 cm$^{-1}$ (where $\alpha > 3\times10^3$ cm$^{-1}$ due to the strong O–H stretching mode of water) was below the detection limit, since the effective path length could not be reduced to below ~3 μm due to the micrometer-scale roughness of the probe surface (~170 μm diameter) as well as the unintentional small tilt of the probe; and 2) no in-plane resolution was demonstrated, due to limited optical throughput, limited hardware control, and lack of probe tapering.

In this work, we overcame the key limitations of our existing endo-FTIR method by fabricating tapered MIR fiber probes with a flat end surface and an aperture size down to ~10 μm, improving the coupling efficiency of the optical setup and the probe, as well as adding the in-plane scanning capability (Figure 1d). The newly established endoscopic micro-FTIR (endo-μFTIR) method provided fine control of path length, with a minimum z below 1 μm, enabling the quantitative determination of $\alpha > 10^4$ cm$^{-1}$ and the spectral coverage across the entire 1000–4000 cm$^{-1}$ range in aqueous solutions. By scanning inside the liquid a few microns above the solid surface, we demonstrated true noncontact chemical mapping with an in-plane resolution down to ~35 μm. Using Beer-Lambert law, we further determined the local thickness of both non-MIR-transparent solid films and liquid regions with a detection limit down to ~17 and 7.3 nm, respectively. This endo-μFTIR method will enable in situ, quantitative, chemically-resolved characterization of many electrochemical, chemical, and biological systems in their native liquid environments.

## 2. RESULTS AND DISCUSSION

### Tapered Probe Microfabrication

To improve the surface flatness and MIR light coupling efficiency, we first polished the input end of the fiber probe. Polishing reduced the arithmetical mean height (Sa) of the fiber cross section from ~17.1 μm to ~1.1 μm (Figure S1a,b). The improved surface flatness resulted in an approximately 55-fold increase in the measured MIR throughput (Figure S1c), likely by reducing surface-scattering losses and thereby improving the coupling efficiency of MIR light into the microprobe.

We further performed microfabrication to reduce the probe aperture diameter from the original 170 μm to 10s of μm and improve the end-surface flatness (Figure 2a). First, the fiber was tapered via chemical etching. Specifically, the probe was partially immersed in a piranha solution covered with a protective layer of pristane oil. During etching, the strongly oxidative piranha solution chemically etched and dissolved the $As_2Se_3$ fiber surface.[59] The accumulation of dissolved reaction products near the fiber surface increased the local solution density and generated a downward convective flow along the fiber. Fresh etchant continuously moved from the oil–etchant interface to the fiber probe, resulting in preferential etching and progressive necking of the probe. Eventually, the thinned neck could no longer support the lower portion of the fiber, which detached and left a tapered probe (Figure 2a). The tapered geometry is dependent on the surface tension at the protective oil–etchant–fiber interface and can possibly be tuned by etchant temperature and concentration.[59,60] A camera was set to monitor the evolution of the tapered shape, and the fiber was taken out once it reached the desired shape. Immediately after etching, the fiber was rinsed with water and isopropyl alcohol (IPA) to remove residual etchant and reaction products. After tapering, the fiber was inserted into the bore of a ceramic ferule and sealed by epoxy using the same method reported in our previous work.[57] To minimize stray light, the tapered fiber was coated with a 10 nm chromium adhesion layer followed by a 100 nm gold layer using the procedure established in our previous work.[57] Focused ion beam (FIB) milling was subsequently used to create the desired probe aperture and improve the flatness of the output end. Since FIB milling exposed the interior of the tapered probe, the resulting output end surface was free of Cr/Au coating except along its perimeter.

Scanning electron microscopy (SEM) images of the probe before and after FIB milling are shown in Figure 2b,c, revealing the overall tapered shape and the FIB-created flat aperture opening. The surface topography of the probe output end was further characterized by optical profilometry, confirming an Sa below 1 μm (Figure S2).

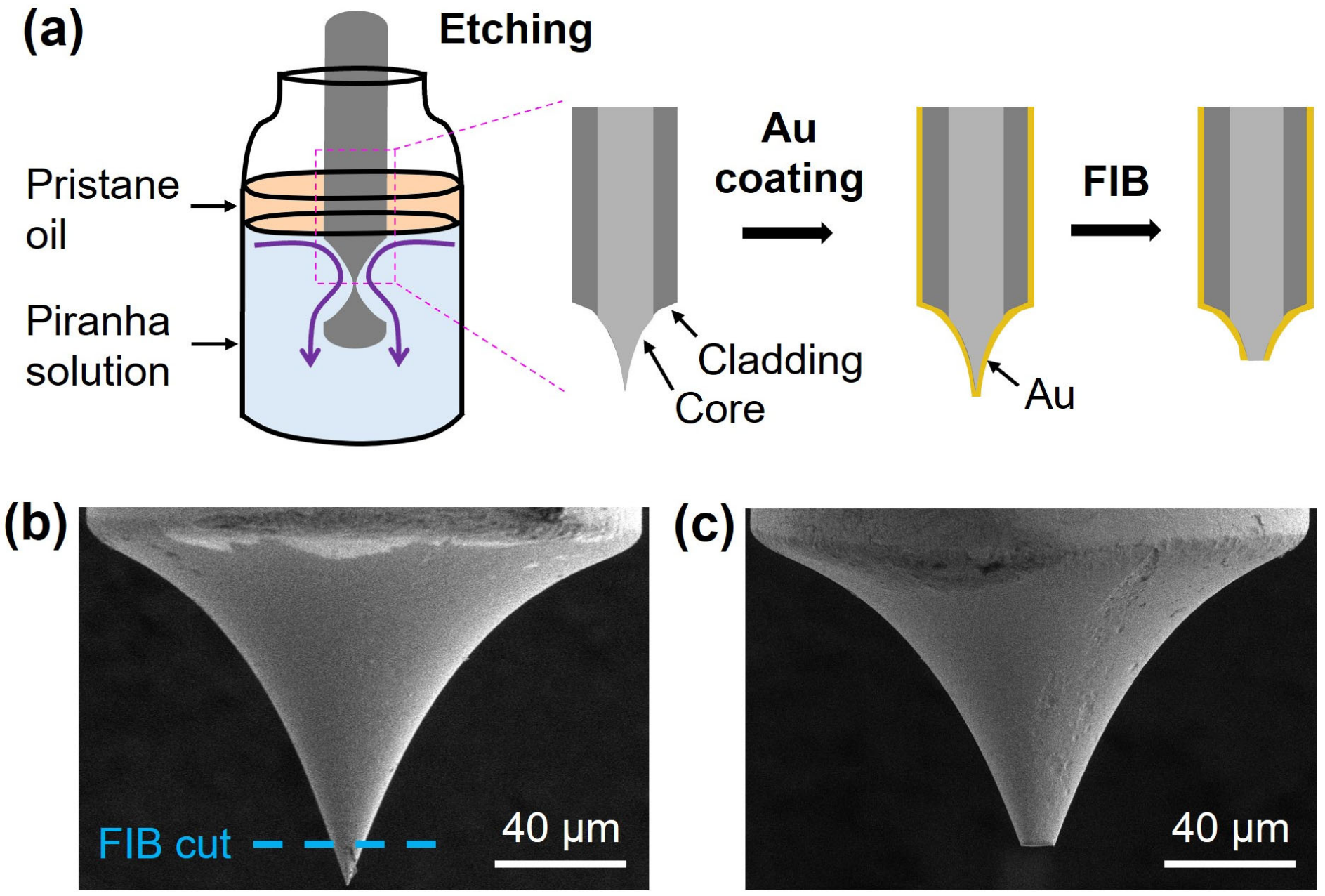

**Figure 2. Microfabrication process of the tapered microprobe.** (a) Schematic of the probe fabrication process. (b,c) SEM images of a microprobe (probe 3) after chemical etching and gold coating (b) and after FIB milling (c). The aperture opening area in (c) is ~111 μm$^2$.

## Tapered Probe Throughput Investigation

Figures 3a and S3 depict the optical configurations used for the experimental MIR measurements and ray-tracing simulations, respectively. The probe geometries used in the simulations were designed to closely reproduce those of the experimentally fabricated probes (Figures S4 and S5). Since the dimensions of the microprobe opening (tens of micrometers) are larger than the MIR wavelength (2.5–10 μm), geometrical optics provides a valid approximation for light propagation. Under this approximation, we performed ray-tracing simulations using a wavelength of 8 μm (wavenumber of 1250 cm$^{-1}$) to predict the optical throughput and beam divergence of the transmitted MIR light (Figure 3b,c). A 12 mm-diameter collimated beam was sent to the optical system and $10^6$ rays were traced. Given the high reflectivity of gold in the MIR region,[61] the gold coating of the tapered probe was modeled as a reflective surface, which is approximated by a sufficiently thick aluminum coating in the simulation. After exiting the probe aperture, the MIR beam gradually diverges, as shown in both the ray distributions in Figure 3b and the cross-sectional irradiance maps in Figure 3c. Using an irradiance threshold of 844 W/cm$^2$ as the noise baseline (Figure S6), we calculated the beam diameter at different z. We found the diameter to be ~42 μm at z = 0 and ~44 μm at z = 18 μm, revealing that the divergence is small within the first ~20 μm. Therefore, we expect the Beer-Lambert law to be applicable within this z range.

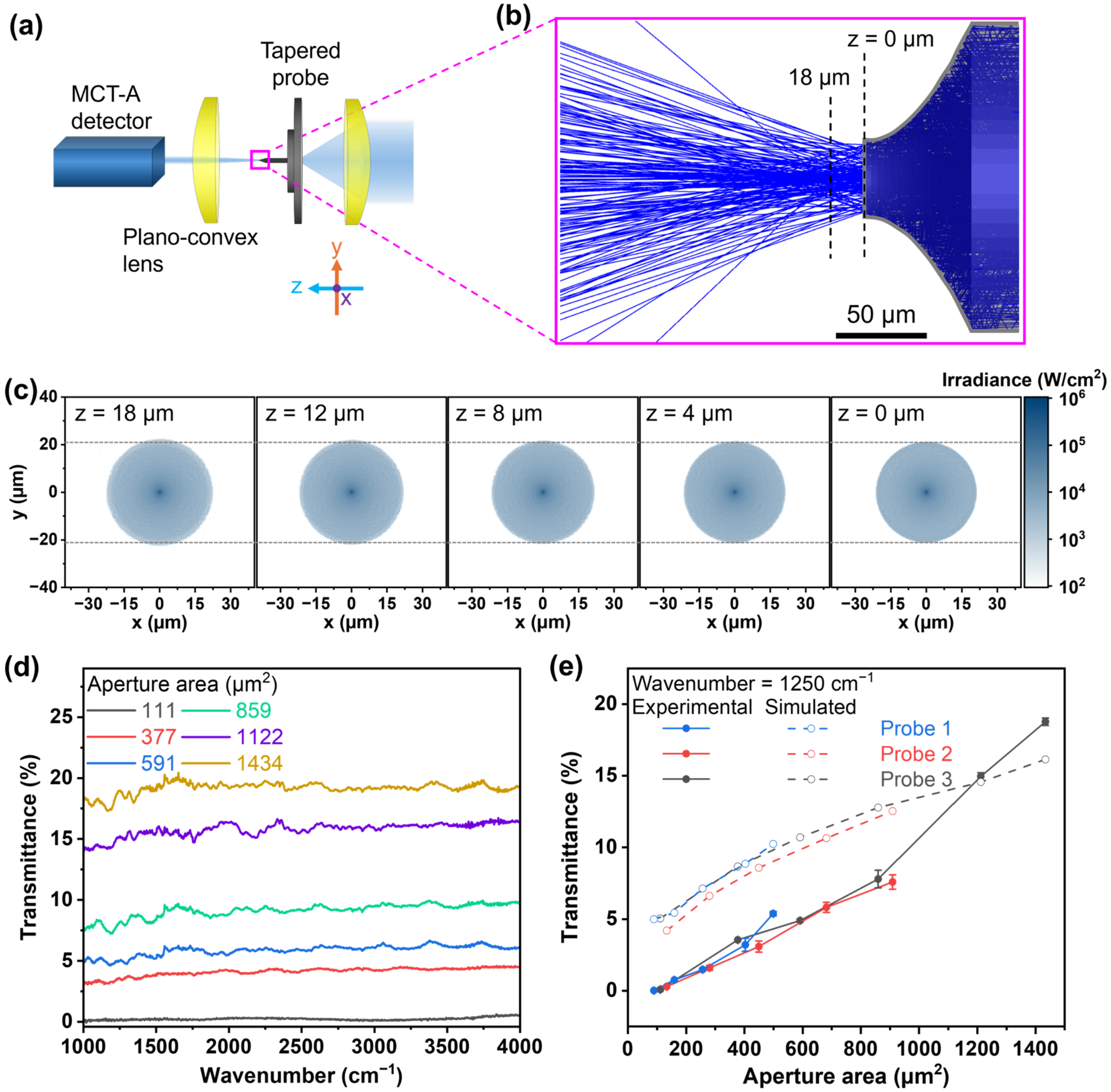

**Figure 3. Tapered probe MIR transmittance quantification.** (a) Schematic of the optical setup used for experimental MIR transmittance measurements. (b) Simulated MIR ray profile of a gold-coated tapered probe (probe 3) with a 42 μm-diameter aperture opening at a wavenumber of 1250 cm$^{-1}$. The overall simulated optical setup is shown in Figure S3, and the geometries of the simulated probes are shown in Figure S5. 900 rays are shown in (b). (c) Irradiance profiles of the outgoing MIR beam cross sections at z = 18, 12, 8, 4, and 0 μm (left to right). z = 0 μm corresponds to the output end of the microprobe. (d) Experimental MIR transmittance spectra of probe 3, measured after progressively cutting the probe to different aperture areas. (e) Experimental (solid circles and solid lines) and simulated (hollow circles and dashed lines) transmittance at multiple aperture areas for three sets of tapered probes at a wavenumber of 1250 $cm^{-1}$. Each experimental data point represents the average value and standard deviation of 2 or 3 separate measurements. SEM images of the three sets of probes are shown in Figure S4.

To experimentally quantify the throughput of the tapered probe as a function of the aperture size, FIB milling was used to progressively increase the probe opening (Figure S4), and the output intensity was measured after each milling step (Figure S7). The throughput was calculated as the ratio of the measured output intensity (Figure S7a,d,g) to the input intensity, and the input intensity was estimated using a 100 μm-diameter pinhole corresponding to the fiber core diameter (Figure S8). Nearly constant transmittance was observed over the broad spectral range of 1000–4000 $cm^{-1}$ for each probe opening, indicating the capability to reliably probe a wide variety of molecular species regardless of the aperture size (Figure 3d). Experimental throughputs obtained from three sets of independently fabricated tapered probes were further compared with those from ray-tracing simulations (Figure 3e). Both the simulated and experimentally obtained transmittance exhibited similar overall trends over the three different taper geometries, showing increased transmittance at larger output apertures. The simulated transmittance was generally higher than the experimental values, which can be attributed to idealizations in the simulation model. Specifically, the simulated probes have simplified geometries with perfectly flat input and output surfaces under ideal optical alignment (Figure S5). In contrast, the experimentally fabricated probes exhibited some irregularities in their geometries (Figure S4) and finite surface roughness (Figure S2), while the optical alignment was performed manually and therefore subject to positioning errors. These deviations from the idealized model can introduce different beam reflection/refraction pathways and/or additional scattering, resulting in lower or different experimental transmittance.

**Precise z-Profiling in Liquid**

Compared to the fiber probes used in our prior endo-FTIR work,[57] the new tapered probes fabricated here feature smaller aperture openings and lower surface roughness (Figure 4a), which are expected to enable more precise path length control down to the sub-micron scale. To test such capabilities, we immersed the fabricated tapered probe in water and performed water z-profiling measurements (Figures 4a and S9). We further corrected the reflection at each interface along the MIR beam path to quantify the MIR transmittance of water (more details in Materials and Methods). Figure 4b shows the water transmittance as a function of wavenumber at a series of z. We observed all the prominent water vibrational modes within the 1000–4000 $cm^{-1}$ spectral range, including H–O–H bending at ~1650 $cm^{-1}$, combination mode (a combination of H–O–H bending and libration) at ~2130 $cm^{-1}$, and O–H stretching at ~2900–3700 $cm^{-1}$.[57]

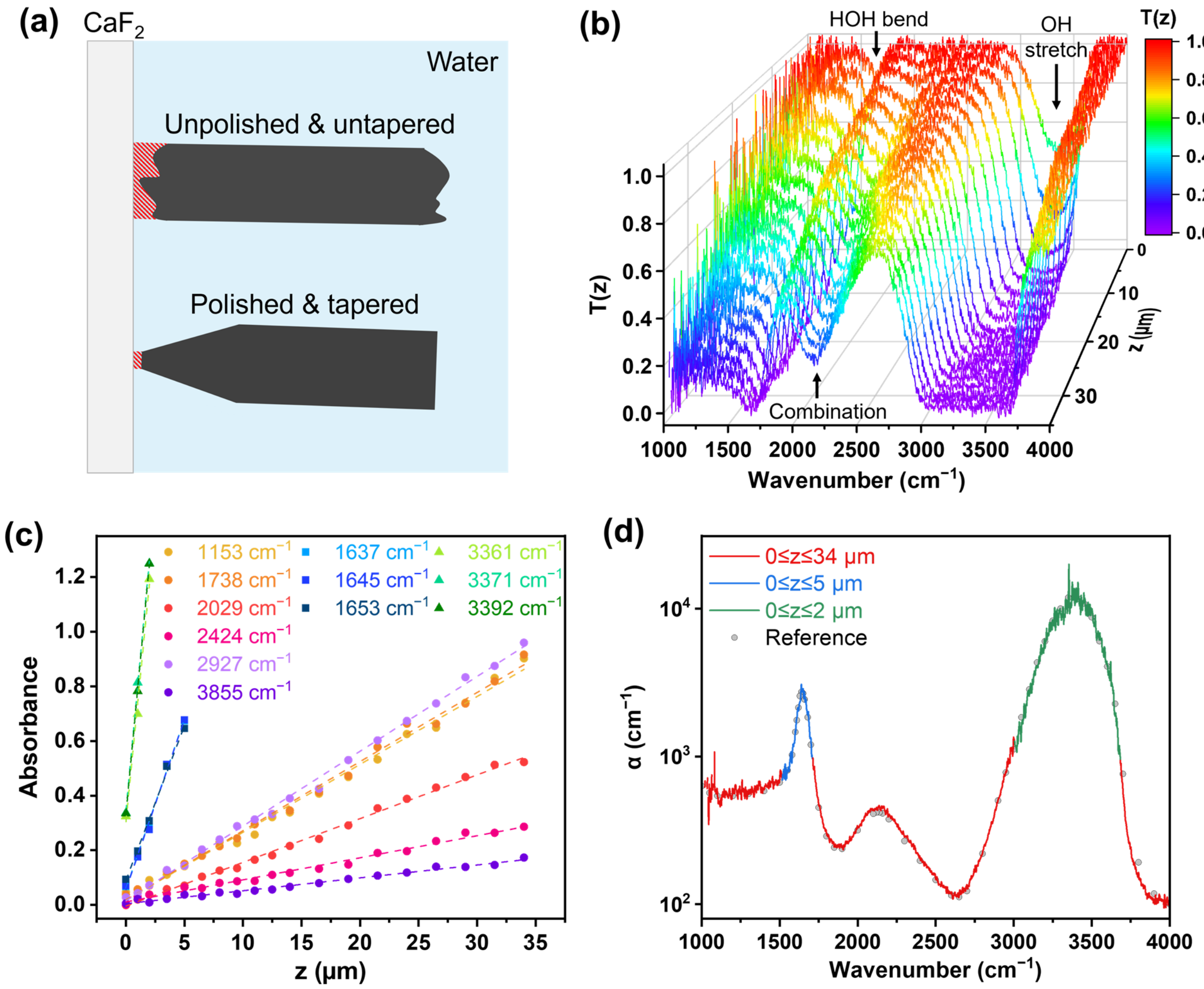

**Figure 4. Water z-profiling with a tapered probe.** (a) Schematic of unpolished and untapered vs polished and tapered probes immersed in water, representing the microprobes used in our prior work[57] and this work, respectively. (b) MIR transmittance of water as a function of wavenumber and z. (c) Absorbance of water at different z, together with their linear fitting, for selected wavenumbers. (d) Absorption coefficient (α) spectrum of water compared with reference. Given the significant variations of the MIR decay length (1/α) across the measured spectral range, α was calculated using different z ranges at different wavenumber ranges to maximize the precision. Specifically, 0≤z≤2 μm was used for wavenumber range 3007–3682 $cm^{-1}$, 0≤z≤5 μm for 1518–1711 $cm^{-1}$, and 0≤z≤34 μm for 1001–1516, 1713–3005, and 3683–4000 $cm^{-1}$. The reference spectrum (gray circles) is adapted with permission from Ref. 62. © 1971 Optical Society of America.

We further converted the MIR transmittance to absorbance, and analyzed its dependence on the path length. As shown in Figure 4c, linear relationships between absorbance and z were observed, confirming the applicability of Beer-Lambert law. We extracted the absorption coefficient (α) of water from the slope of the linear fits of absorbance vs z at each wavenumber, and plotted the resulting spectrum in Figure 4d. Since the MIR decay length (1/α) varies strongly with wavenumber, different z ranges were chosen for the linear fitting of absorbance at different wavenumber ranges, in order to obtain the precise α values. The obtained α values across the 1000–4000 $cm^{-1}$ range are all in good agreement with existing standard results obtained from wedge-shaped transmission cells.[62] Note that these wedge-shaped cells do not provide spatial resolution and are therefore limited to spatially homogeneous samples, in contrast to our endo-μFTIR method, which enables quantitative characterization of spatial heterogeneities along both the z and in-plane directions.

The highest α value we quantified was ~$1.2\times10^4$ $cm^{-1}$ at the peak of the O–H stretching band, corresponding to an accessible minimum MIR decay length of ~0.8 μm. This is a notable improvement from our previous work,[57] where the complete O–H stretching band could not be resolved due to the residual water trapped between the rough probe surface and the substrate (Figure 4a). The larger accessible ranges of α and z were achieved in our current endo-μFTIR setup via multiple improvements, including: 1) increased signal-to-noise ratio by changing the objective lens used in our previous report[57] to a plano–convex lens with larger clear aperture (Figure S10); 2) mechanical polishing of the input end of the probe (Figure S1); 3) use of a tapered probe with a smaller aperture, which reduced the residual water thickness when the probe was unintentionally tilted by a small angle (such uncontrolled small tilts are currently unavoidable in all existing scanning probe methods; Figure 4a); and 4) reduced roughness of the tapered probe surface at the output end (Figure S2b), which further minimized the residual water volume at low z.

**In-Plane Chemical and Morphological Mapping of Solids and Near-Surface Liquids**

To benchmark the in-plane chemical and morphological mapping capabilities, we fabricated a micropatterned SU-8 resolution target consisting of alternating SU-8 stripes and exposed $CaF_2$ regions (Figure S11). A tapered microprobe (probe 3) with an opening diameter of 38.4 to 45.4 μm (Figures 5a and S12) was used to image the fabricated resolution target in water (Figure 5b). The stripe width and separation (exposed $CaF_2$ regions) are both ~200 μm, much larger than the ~40 μm aperture diameter of the microprobe. At the current stage, precise position and feedback control have not yet been implemented. Nevertheless, the existing setup already enables precise quantification of all MIR-absorbing substances along the optical path, as long as the microprobe is sufficiently close to the region of interest. MIR absorbance spectra were acquired during in-plane scanning along the x direction, which is perpendicular to the stripe direction of the resolution target. We scanned over a distance of ~650 μm, sufficient to resolve more than 1.5 periods of the resolution target pattern. During in-plane scanning, we observed periodic appearances of the peaks at 1246 $cm^{-1}$ and 1508 $cm^{-1}$, due to the C−O−C asymmetric stretch and the aromatic C−C stretch (in-ring) of SU-8, respectively[63,64] (Figures 5c and S13). The water H–O–H bending (~1650 $cm^{-1}$) and combination (~2130 $cm^{-1}$) bands also varied during scanning, reflecting changes in the local water path length above the patterned surface.

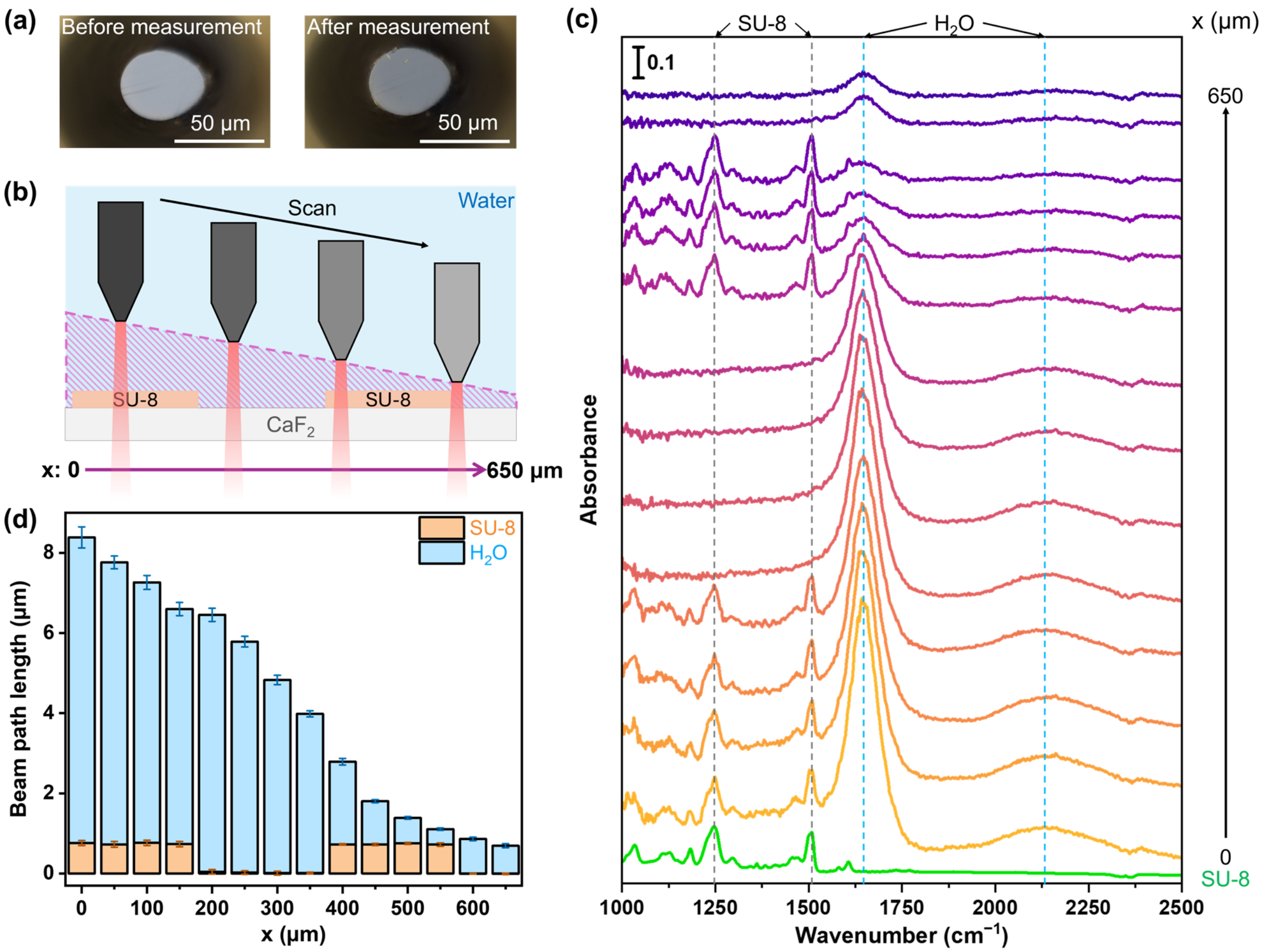

**Figure 5. In-plane chemical and morphological mapping of a resolution target in water using endo-μFTIR.** (a) Optical microscope images of the tapered probe before and after endo-μFTIR measurements. (b) Side-view schematic illustrating the in-plane endo-μFTIR measurements of the micropatterned resolution target in water. Purple dashed regions illustrate the active probing volume, which includes both the SU-8 micro-stripes and the surrounding water. (c) MIR absorbance spectra of the resolution target in water obtained during the scan along the x direction with a step size of 50 μm (orange to purple), as well as a reference SU-8 spectrum in air (green). The SU-8 reference spectrum was obtained by measuring the absorbance of a continuous SU-8 film on $CaF_2$ substrate using a pinhole configuration in air, with the bare $CaF_2$ background subtracted (Figure S15a). (d) Quantification of SU-8 and water thickness as a function of x. SU-8 thickness was calculated from 9 spectral data points ranging from 1500–1516 cm$^{-1}$. Water thickness was calculated from 17 data points ranging from 1630–1660 cm$^{-1}$. Error bars represent the standard deviation.

To quantitatively deconvolute the absorbance spectra and extract the local SU-8 and water layer thickness, the absorption coefficient values of each component across the whole spectral range are required. Water absorption coefficient spectrum was already determined in Figure 4d. We further obtained the absorption coefficient of SU-8 ($\alpha_{SU8}$) by measuring the absorbance spectrum of a continuous SU-8 film with predetermined thickness, and applying the Beer-Lambert law (Figures S14 and S15):

$$\alpha = \frac{A_{exp} ln(10)}{z}, \quad (1)$$

where $A_{exp}$ is the experimentally measured MIR absorbance.

Based on the raw absorbance spectra in Figure 5c, we further subtracted water contribution to isolate the SU-8 absorbance (Materials and Methods; Figure S16). Using the experimentally determined $\alpha_{SU8}$, eq 1 was applied to the water-subtracted SU-8 absorbance to calculate the local SU-8 thickness. Since SU-8 exhibited two pronounced peaks (Figures 5c and S16b), both can be used to extract the local thickness of the SU-8 film. The thickness profiles calculated using the 1500–1516 $cm^{-1}$ and 1238–1254 $cm^{-1}$ spectral range are shown in Figure 5d and Figure S17, respectively. In both results, we observed SU-8 stripes with nearly constant thickness, ~200 µm width and ~200 µm separation. An average thickness of 0.74 ± 0.02 µm and 0.74 ± 0.08 µm were obtained from Figure 5d and Figure S17, respectively. These values are close to the thickness of 0.75 ± 0.15 µm measured by AFM (Figure S14). Similarly, water absorbance was isolated by subtracting the SU-8 contribution from the spectra acquired during in-plane scanning (Materials and Methods; Figure S16a). The local water thickness was quantified from the H–O–H bending mode (1630–1660 $cm^{-1}$) using eq 1. As shown in Figure 5d, we observed a decrease in the local water thickness from left to right (i.e., with increasing x) within both the individual SU-8 and $CaF_2$ substrate regions, due to the roughly downward probe scanning trajectory (Figure 5b). However, across the right boundaries of each SU-8 stripe, water thickness suddenly increased (from 5.87 ± 0.17 to 6.41 ± 0.16 µm as x increased from 150 to 200 µm, and from 0.38 ± 0.03 to 0.86 ± 0.04 µm as x increased from 550 to 600 µm), since the transition from SU-8 to $CaF_2$ resulted in increased path length through water. The sum of the SU-8 and water thickness decreased from ~8.39 ± 0.27 to ~0.68 ± 0.05 µm as x increased from 0 to 650 µm (Figure 5d), again reflecting the downward scanning trajectory of the microprobe (Figure 5b).

The finite water thickness measured throughout the scan (Figure 5d), together with the unchanged probe surface after the measurement (Figure 5a), indicated that the probe was not in contact with the solid surface during scanning.

To evaluate the spatial resolution of the tapered probe, we conducted endo-µFTIR measurement across an edge of the SU-8 stripe pattern (still immersed in water) using step sizes smaller than the aperture size of the probe (Figure 6a). The same tapered probe shown in Figure 5a with an aperture size of ~40 µm was used. When the probe scanned on top of the SU-8 region (from x = 0 to 125 µm), the SU-8 absorbance (peaks at 1246 $cm^{-1}$ and 1508 $cm^{-1}$) remained mostly constant while the water absorbance (peaks at 1650 $cm^{-1}$ and ~2130 $cm^{-1}$) continuously increased (Figures 6b and S18), consistent with the constant thickness of the SU-8 film and the upward probe scanning trajectory (Figure 6a). As the probe moved from the SU-8 region toward the exposed $CaF_2$ region, the SU-8 absorbance decreased sharply from x = 125 to 160 µm (Figures 6b and S18), consistent with the local composition change of the solid. Using the same methods described for Figure 5, we further extracted the quantitative water and SU-8 thickness across the scanned region (Figures

S19 and 6c). The SU-8 thickness extracted from the 1500–1516 $cm^{-1}$ region was 0.64 ± 0.01 μm at 0 ≤ x ≤ 125 μm, which decreased sharply to zero as x increased from 125 μm to 160 μm (Figure 6c). Thickness quantification using the 1238–1254 $cm^{-1}$ region yielded similar results (Figure S20). The ~35 μm width of this transition region (x = 125–160 μm; red dashed lines in Figure 6c) therefore provides an experimental estimate of the lateral spatial resolution. While the aperture diameter of the used probe was 38.4–45.4 μm (Figure 5a), our ray-tracing simulation revealed that the photon flux is stronger at the center of the probe and significantly weakens near the edge of the probe (Figure S6). As such, it is reasonable to achieve an in-plane spatial resolution slightly finer than the probe opening.

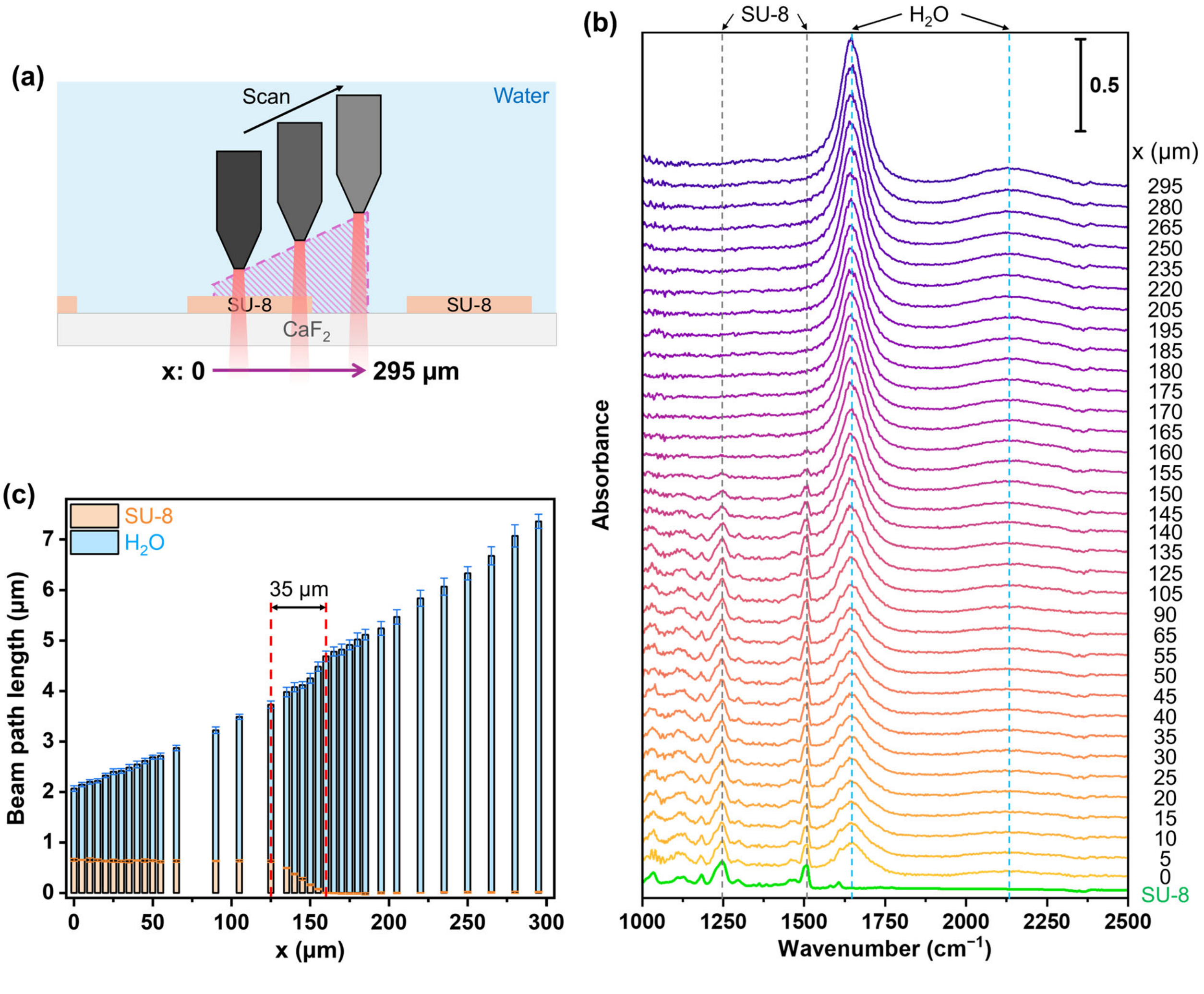

**Figure 6. Fine scan and resolution determination of endo-μFTIR.** (a) Side-view schematic illustrating endo-μFTIR measurement across an edge of an SU-8 stripe in water. (b) MIR absorbance obtained as the microprobe scanned along the x direction from 0 to 295 μm (orange to purple), as illustrated in (a). SU-8 spectrum in air (green), same as that in Figure 5c, is also shown as a reference. (c) Quantification of local SU-8 and water thickness as a function of x, obtained from the data in (b). SU-8 thickness was calculated from 9 spectral data points ranging from 1500–1516 $cm^{-1}$, while water thickness was calculated from 17 data points over 1630–1660 $cm^{-1}$. Red

dashed vertical lines mark the transition from SU-8 to the $CaF_2$ substrate, revealing an in-plane resolution of 35 μm.

We further quantified the z detection limit of endo-μFTIR for both SU-8 and water, defined here as the minimum detectable optical path length. The z detection limit was calculated based on the noise background of the instrument, the absorption coefficient (α) of the probed material, and Beer-Lambert law (eq 1) (Materials and Methods). The calculated detection limits were ~17 nm for SU-8 (using the 1246 $cm^{-1}$ vibrational peak) and ~7.3 nm for water (using the 3390 $cm^{-1}$ peak). These results demonstrate that endo-μFTIR is sensitive to nanometer-scale changes in optical pathlength along the z direction, enabling the detection of subtle variations in thickness for both solid and liquid samples.

## 3. CONCLUSIONS

By integrating a tapered fiber microprobe and a 3D scanning stage in a transmission-based MIR optical setup, we developed the endo-μFTIR method. This technique enabled truly noncontact, quantitative chemical and morphological mapping of solid structures and their surrounding liquid environment, achieving ~35 μm in-plane resolution. When applied to an SU-8 resolution target in water, we achieved a vertical detection limit of 17 nm for SU-8 and 7.3 nm for water. We expect this developed technique to be highly promising for applications in many chemical, electrochemical, and biological systems where heterogeneous solid–liquid interfaces are present, such as batteries, catalytic processes/systems, living cells, and more.[4–6,9–11]

Future developments of the endo-μFTIR method will focus on two improvements. One is to incorporate a feedback-control system for active regulation of the probe–sample distance during in-plane or 3D scanning. The MIR absorbance of the liquid, as a direct indicator of local liquid thickness (probe–substrate distance), can serve as the input for position feedback. Such feedback control would enable high-speed chemical imaging while avoiding probe–substrate contact. Another future improvement is to replace the FTIR light source (globar light) by an MIR laser with much stronger output power, to enable significantly higher spatial resolution (e.g, sub-micron in-plane resolution and sub-nanometer z detection limit) and higher chemical sensitivity. For example, quantum cascade lasers (QCLs) offer a spectral power density 4–5 orders of magnitude higher than that of the FTIR light source.[65] Accordingly, under otherwise identical probe and sample conditions, the use of QCLs is expected to provide a substantially higher signal-to-noise ratio than FTIR-based measurements.

## 4. METHODS

### Fiber Input and Output End Processing

As reported in our previous study, the bulk $As_2Se_3$ optical fiber (IRflex Corporation, IRF-Se-100R) were cut into ~30–35 mm segments, followed by removal of the polyacrylate (PA) coating.[57] After

PA removal, the fiber was inserted into a steel needle tube and heated at 140 °C for ~2 hours for straightening. The fiber was then placed into a ferrule (Thorlabs, CFX230-10) and secured on the funnel end with carbon paint (Ted Pella DAG-T-502). The end of the fiber was allowed to extend less than ~1 mm from the flat end of the ferrule. The ferrule containing fiber was then inserted into a polishing puck (Thorlabs, D50-F) with the flat side facing down. Wax was melted and used to secure the fiber on the flat side in the puck. The fiber input end on the flat side of the ferrule was polished to reduce the surface roughness using a series of alumina lapping sheets (Thorlabs, LF5P, LF3P, LF1P, and LF03P). The size and flatness of the fiber output end were precisely engineered using FIB milling (FEI Helios 600i Dual Beam SEM/FIB).

**Etchant Preparation for Chemical Etching of Tapered Probes**

A piranha solution was prepared by first heating a water bath to 50 °C. In a separate water bath for cooling, 3.5 mL of sulfuric acid (95.0–98.0%, Fisher Chemicals, A300S-212) was added to a 20 mL glass vial, followed by the dropwise addition of 1.5 mL of 27% $H_2O_2$ solution (prepared from 30% $H_2O_2$, Fisher Chemicals, H325-500). The mixture was stirred gently until gas evolution ceased. Subsequently, 1 mL of pristane oil (synthetic, ≥98%, Sigma-Aldrich, P2870-100ML) was carefully dropped on top of the piranha solution to serve as a protective solvent layer.

**Ray-Tracing Simulation and Transmittance Calculation**

Ray-tracing simulations were performed using Zemax OpticStudio to investigate light propagation through the tapered $As_2Se_3$ fiber probe. The simulation configuration was constructed based on the experimental setup (Figures S3b and S21), and the probe geometries were designed to closely reproduce those of the experimentally fabricated tapered probes (Figures S4 and S5). A 12 mm-diameter collimated beam with a wavelength of 8 μm was incident on a plano-convex lens (Thorlabs, LA 7656-G, 50.0 mm focal length (FL)) and $10^6$ rays were traced through the system in non-sequential mode. The refractive indices of the fiber core and cladding were set to 2.72 and 2.70, respectively. The gold-coated outer surface of the tapered probe was modeled as a reflective surface. To account for potential alignment errors in the experiment, a 0.5 mm leftward offset was introduced between the focal point of the incident beam and the fiber input surface in the simulation.

The optical throughput of each microprobe was calculated as the ratio of the output optical power to the input optical power. The input optical power was recorded using a 100 μm × 100 μm detector positioned at the fiber input end (Figure S3a), whereas the output power was recorded using a 200 μm × 200 μm detector positioned at multiple z positions relative to the microprobe output end (Figure S3b).

**Fabrication of the SU-8 Resolution Target**

To demonstrate the in-plane scanning capability of the endo-μFTIR technique, we made a resolution target composed of SU-8 photoresist (Kayaku Advanced Materials, TF 6000.5) stripes with a width of 200 μm on a $CaF_2$ substrate (International Crystal Laboratory, 0002D-1457). We spin-coated SU-8 on a $CaF_2$ substrate at the speed of 1000 rpm and soft baked it at 110 °C. The sample was then exposed to a visible light at a wavelength of 375 nm with an exposure energy of 100 mJ/cm$^2$ using a digital mask (Heidelberg MLA150 aligner). We then baked the sample for another 2 minutes at 110 °C and developed it using the SU-8 developer. The resolution target was imaged by optical microscope (Keyence VK-X1000 3D Laser Scanning Confocal Microscope) and AFM (Asylum Cypher S).

**FTIR Measurements**

FTIR measurements were conducted based on our setup that was previously built in the transmission chamber of a commercial FTIR spectrometer (Thermo Nicolet iS50) (Figure S21),[57] which has a Polaris MIR source with a spectral range of 10–9600 cm$^{-1}$ and a liquid nitrogen-cooled MCT-A detector sensitive to the spectral range of 600–11700 cm$^{-1}$. A ZnSe plano convex lens (Thorlabs, LA7656-E4, 50.1 mm FL, >20.3 mm clear aperture) was used to focus the MIR light onto the tapered fiber (Figure S10a). The incoming MIR beam (incident on the plano convex lens) has a diameter of ~12 mm. Compared with the objective lens used in our previous work (Edmund Optics, 88-448, 0.08 numerical aperture (NA), 18 mm FL, 5 mm clear aperture), the ZnSe plano-convex lens provides a larger clear aperture, allowing a greater fraction of MIR light to be collected and coupled into the fiber. This modification increased the measured MIR throughput by ~1.7 times (Figure S10b).

The SU-8/$CaF_2$ resolution target was mounted on an XYZ translation stage (Thorlabs, MT3A), providing 3D positional control with ~1 μm precision. After the MIR light was transmitted through the tapered fiber and the sample, it was refocused through another ZnSe plano–convex lens (Thorlabs, LA7656-E4, 50.1 mm FL) before reaching the detector.

For precise z profiling measurement in liquid, z = 0 μm was defined as the z position below which negligible changes in the raw intensity can be observed. All FTIR measurements were conducted by setting IR resolution = 4 cm$^{-1}$, data spacing = 1.928 cm$^{-1}$, number of scans = 16, gain = 8, optical velocity of the moving mirror = 0.1581 cm/s, and beam aperture = 230.

**Atomic Force Microscopy Imaging**

Samples were mounted onto steel discs using double-sided Kapton tape prior to imaging. AFM measurements were performed in ambient air using a Cypher S AFM (Asylum Research, Oxford Instruments) operated in tapping (AC) mode with piezo acoustic excitation. Two probe types were used: FS-1500AuD probes (Asylum Research, Oxford Instruments), with a manufacturer-specified spring constant of 6–26 N/m and nominal resonance frequency of ≈1.5 MHz, and PPP-NCHAuD probes (NanoAndMore USA), with a manufacturer-specified spring constant of 42 N/m and a measured resonance frequency of ≈287 kHz.[66–68] Both probe types had nominal tip radii below 10 nm according to manufacturer specifications. The free oscillation amplitude was typically set to either ≈800 mV or 1 V, with the amplitude setpoint maintained at approximately 65–75% of the free amplitude. Images were acquired at a scan rate of 0.5 Hz with a resolution of 256 × 256 pixels over scan areas of either 20 × 20 μm or 30 × 30 μm. As the measurements focused on micron-scale feature heights rather than nanoscale surface morphology, probe selection was not expected to significantly influence the results; accordingly, consistent height measurements were obtained using both probe types. Image processing and analysis were performed using Gwyddion.

**Reflectance Correction**

When light propagates through multiple interfaces, reflections at each interface contribute to the measured raw intensity (Figure S22). To apply Beer-Lambert law correctly, reflection losses at interfaces needed to be corrected.[22] For the measurements of $CaF_2$ substrate in air, $CaF_2$ substrate in water, and $CaF_2$ substrate/SU-8 in water (Figure S22), the corresponding raw intensities can be written as:

$$I_{air} = I_0(1-\eta_{others})exp(-\alpha_{air}z_{air})(1-R_{CaF_2/air})(1-R_{air/probe}), \quad (2)$$

$$I_{H_2O} = I_0(1-\eta_{others})exp(-\alpha_{H_2O}z_{H_2O})(1-R_{CaF_2/H_2O})(1-R_{H_2O/probe}), \quad (3)$$

$$I_{SU8+H_2O} = I_0(1-\eta_{others})exp(-\alpha_{H_2O}z_{H_2O}-\alpha_{SU8}z_{SU8})(1-R_{CaF_2/SU8})$$
$$(1-R_{SU8/H_2O})(1-R_{H_2O/probe}). \quad (4)$$

where $I_{air}$, $I_{H_2O}$, and $I_{SU8+H_2O}$ are the raw measured intensities measured for $CaF_2$ substrate in air, $CaF_2$ substrate in water, and $CaF_2$ substrate/SU-8 in water configurations, respectively. The quantities $\alpha_{air}$, $\alpha_{H_2O}$, and $\alpha_{SU8}$ are the absorption coefficients of air, $H_2O$, and SU-8, respectively, while $z_{air}$, $z_{SU8}$, and $z_{H_2O}$ are their corresponding optical path lengths. $R_{CaF_2/air}$, $R_{air/probe}$, $R_{CaF_2/H_2O}$, $R_{H_2O/probe}$, $R_{CaF_2/SU8}$, $R_{SU8/H_2O}$, and $R_{H_2O/probe}$ are the reflectance between $CaF_2$ substrate/air, air/probe, $CaF_2$ substrate/$H_2O$, $H_2O$/probe, $CaF_2$ substrate/SU8, SU8/$H_2O$, and $H_2O$/probe interface, respectively. $I_0$ is the raw intensity of the FTIR output beam. $\eta_{others}$ corresponds to the total losses of the MIR beam before reaching the microprobe output and after reaching the $CaF_2$ substrate surface; these losses were assumed to be the same for the $CaF_2$ in air, $CaF_2$ in water, and $CaF_2$/SU-8 in water configurations, since identical optical setup was used outside the probe and sample.

Reflectance ($R$) at each interface was calculated using the Fresnel Equations, assuming normal incidence to the interface:[69]

$$R = \frac{(n_1-n_2)^2+(k_1-k_2)^2}{(n_1+n_2)^2+(k_1+k_2)^2}, \quad (5)$$

where $n_1$ and $n_2$ are the real parts of the refractive indices of media 1 and 2, respectively, and $k_1$ and $k_2$ are the corresponding imaginary parts. The real refractive indices of probe ($n_{probe} \approx$ 2.7, as specified by the manufacturer) and air ($n_{air} \approx 1.0$) were assumed to be constant over the MIR spectral range, with their imaginary components neglected. For $CaF_2$ substrate, the wavelength-dependence of the real refractive index was considered,[70] while its imaginary component was neglected. For $H_2O$ and SU-8, both the real and imaginary components of the refractive index were considered to be wavelength dependent.[30,71] The resulting wavelength-dependent reflectances at multiple interfaces are shown in Figure S23. Among these interfaces, reflection losses were dominated by the $H_2O$/probe and air/probe interfaces, and these contributions were therefore included in the transmittance correction. For spectra of pure water, the corrected transmittance is given by:

$$\frac{I_{H_2O}}{I_{air}} = \frac{1 - R_{H_2O/probe}}{1 - R_{air/probe}} exp\left(-\alpha_{H_2O} z_{H_2O}\right). \tag{6}$$

For spectra of SU-8 + water, the corrected transmittance is given by:

$$\frac{I_{SU8+H_2O}}{I_{air}} = \frac{1 - R_{H_2O/probe}}{1 - R_{air/probe}} exp\left(-\alpha_{H_2O} z_{H_2O} - \alpha_{SU8} z_{SU8}\right). \tag{7}$$

Here, $exp(-\alpha_{air} z) = 1$, since $I_{air}$ was observed to be independent of z (indicating $\alpha_{air}$ is negligible), as confirmed in our previous work.[57]

**SU-8 and Water Separation Methods**

SU-8 background removal and $H_2O$ thickness determination

For each spectrum containing a contribution from SU-8, the absorbance difference between the SU-8 peak at 1246 $cm^{-1}$ and the reference point at 1198 $cm^{-1}$ was first calculated. The reference SU-8 spectrum measured in air was then scaled to match this absorbance difference and subtracted from the measured spectrum of SU-8 + $H_2O$ to remove the SU-8 contribution. After SU-8 subtraction, the water thickness for each spectrum was calculated using the reference absorption coefficient of $H_2O$ according to:

$$z_{water} = \frac{ln(10) A_{water}}{\alpha_{water\ reference}}, \tag{8}$$

where $A_{water}$ is the IR absorbance after subtraction of the SU-8 contribution, $z_{water}$ is the calculated water thickness for each spectrum, and $\alpha_{water\ reference}$ is the reference absorption coefficient of water.[30]

$H_2O$ background removal and SU-8 thickness determination

The $H_2O$ thickness in each spectrum was first determined using the method described above. Based on the obtained $H_2O$ thickness, the corresponding water background absorbance was calculated and subtracted from each spectrum to isolate the SU-8 contribution. The SU-8 thickness was then quantified using the reference SU-8 absorption coefficient.

$$A_{water\ background} = \alpha_{water\ avg} \frac{z_{water}}{ln(10)}, \tag{9}$$

$$A_{SU8} = A - A_{water\ background}, \tag{10}$$

$$z_{SU8} = \frac{ln(10) A_{SU8}}{\alpha_{SU8}}, \tag{11}$$

where $\alpha_{water\ avg}$ is the average water absorption coefficient obtained from water-only spectra acquired at multiple x positions, with the absorption coefficient at each position determined using the Beer-Lambert law. $z_{water}$ is the water thickness determined for each spectrum from eq 8, and $A_{water\ background}$ is the corresponding calculated water background absorbance spectrum. $A$

represents the measured absorbance spectrum at each x position. After subtraction of the water background, the remaining absorbance ($A_{SU8}$) was attributed to SU-8. The corresponding SU-8 thickness ($z_{SU8}$) was then calculated using the SU-8 absorption coefficient ($\alpha_{SU8}$) determined in Figure S15b. All absorbance spectra used in eqs 9–11 were corrected for reflection losses prior to the analysis.

**z Detection Limit Determination**

The detection limit of the absorbance was defined as:[72]

$$A_{min} = \mu + 3\sigma, \tag{12}$$

where $\mu$ and $\sigma$ represent the mean and standard deviation of the instrument noise baseline, respectively, calculated from the wavenumber range of 1842–3242 cm$^{-1}$. The instrument noise baseline spectrum was obtained with the probe fully immersed in water on a $CaF_2$ substrate at a sufficiently large z, where the MIR signal was completely attenuated by water such that the remaining signal was dominated by the instrumental noise (Figure S24). The z detection limit was calculated from the detection limit of the absorbance and the experimentally determined absorption coefficient using the Beer-Lambert law (eq 1).

## ASSOCIATED CONTENT

### Supporting Information

The Supporting Information is available free of charge online.

Throughput improvement resulted from polishing the fiber input side, optical profilometry measurements of the input and output ends of probe 3, ray-tracing simulation setup, images of microprobes before and after a series of FIB milling, microprobe geometries with different shapes and aperture openings used in the ray-tracing simulation, beam divergence evaluation using ray tracing simulation, experimental MIR throughput of three sets of independently fabricated tapered probes, raw intensity spectrum through a 100 μm-diameter pinhole, raw intensity spectrum for water z-profiling via endo-μFTIR, throughput improvement by replacing the objective lens to plano-convex lens, optical images of the resolution target before and after endo-μFTIR in-plane scanning measurements, side-view optical image of the microprobe used to produce the results in Figures 5 and 6, overlaid MIR absorbance spectra obtained during the resolution target measurement in Figure 5, AFM characterization to determine the height of the SU-8 film, SU-8 absorption coefficient determination, separated absorbance spectra of water and SU-8 extracted from Figure 5c, quantification of SU-8 and water thickness during in-plane scanning from x = 0 to 650 μm, overlaid MIR absorbance spectra obtained during the resolution target measurement in Figure 6, separated spectra of water and SU-8 during the scan along x direction over an edge of an SU-8 stripe in water, quantification of SU-8 and water thickness during in-plane scanning from x=0 to 295 μm, experimental setup of endo-μFTIR, schematics illustrating reflection losses in the optical configurations used for measurements in air, water, and SU-8 + water, reflectance as a function of wavenumber at multiple interfaces, and instrument noise baseline (PDF)

**Notes**

The authors declare no competing financial interest.

## ACKNOWLEDGMENTS

We acknowledge the support from the Beckman Young Investigator Award provided by the Arnold and Mabel Beckman Foundation and the Sloan Research Fellowship from the Alfred P. Sloan Foundation. J.K. acknowledges financial support from the TechnipFMC Educational Fund at the University of Illinois Urbana-Champaign. The experiments were performed in part in the Materials Research Laboratory at the University of Illinois Urbana-Champaign.